\documentclass[conference]{IEEEtran}

\usepackage{graphics,graphicx} 

\usepackage{amsmath}
\usepackage{amssymb}
\usepackage{tikz}
\usepackage{amsthm}
\usepackage{comment}
\usepackage[export]{adjustbox}
\usepackage{mathcomSTEP}
\usepackage{subcaption}
\usepackage{setspace}	
\usepackage[utf8]{inputenc}
\usepackage{cite}
\usepackage{float}
\usepackage{lipsum}
\usepackage{stfloats}

\usetikzlibrary{positioning}
\allowdisplaybreaks

\newtheorem{prop}{Proposition}

\newtheorem{theorem}{Theorem}
\theoremstyle{remark}
\newtheorem{rem}[thm]{\bf Remark}

\newtheorem{definition}[thm]{\bf Definition}

\usepackage{tikz}
\usetikzlibrary{positioning}
\usetikzlibrary{arrows,fit}
\usetikzlibrary{shapes.geometric}
\usetikzlibrary{positioning,calc}
\usetikzlibrary{decorations.markings,shapes,arrows}

\usepackage{epstopdf}
\usepackage{pgfplots,setspace}
\usepackage{circuitikz}

\tikzstyle{int}=[draw, fill=blue!10, minimum height = .5 cm, minimum width=1 cm,thick ]
\tikzstyle{int1}=[draw,  minimum height = .15 cm, minimum width=1 cm,thick ]
\tikzstyle{sum}=[circle, fill=blue!10, draw=black ]

\newcommand{\Wt}{\widetilde{W}}
\newcommand{\Nrx}{N_{\rm r} }
\newcommand{\Ntx}{N_{\rm t} }
\newcommand{\Ntq}{N_{\rm tq} }
\newcommand{\rank}{\textsf{rank}}

\allowdisplaybreaks

\newcounter{image}
\pgfmathtruncatemacro{\recordwidth}{1}
\pgfmathtruncatemacro{\recordheight}{1}

\usepackage{xparse}

\tikzset{drawinside/.code args={#1}{%
           \draw[dashed]($(#1.west)!0.3!(#1.center)$)--($(#1.east)!0.3!(#1.center)$);
           \draw($(#1.south west)!0.3!(#1.west)!0.2!(#1.center)$)-|(#1.center);
           \draw (#1.center)|- ($(#1.north east)!0.3!(#1.east)!0.2!(#1.center)$);
       }
}

\tikzset{record/.style args={#1 and #2}{
        rectangle,draw,minimum width=#1, minimum height=#2
    }
}

\NewDocumentCommand{\drawrecord}{d()}{
\stepcounter{image}
\IfNoValueTF{#1}{
\node[record=\recordwidth cm and \recordheight cm,name=a\theimage]{};
}
{
\node[record=\recordwidth cm and \recordheight cm,name=a\theimage]at(#1){};

}
\node[drawinside={a\theimage}]{};
}

\newcommand{\rr}{\mathrm {r}}
\newcommand{\ror}{\mathrm {r_{\zerov}}}
\newcommand{\rpar}{\mathrm {r_{p}}}
\newcommand{\rssps}{\mathrm {r_{ssps}}}

\begin{document}
	
	\title{
		On MIMO  Channel Capacity\\ with  Output Quantization Constraints
	}
	\author{
		
		\author{
			\IEEEauthorblockN{Abbas Khalili}
			\IEEEauthorblockA{\textit{NYU } \\
				New York, USA \\
				ako274@nyu.edu}
			\and
			\IEEEauthorblockN{Stefano Rini}
			\IEEEauthorblockA{\textit{NCTU} \\
				{Hsinchu, Taiwan}\\
				stefano@nctu.edu.tw
			}
			\and
			\IEEEauthorblockN{Luca Barletta}
			\IEEEauthorblockA{\textit{Politecnico di Milano} \\
				Milano, Italy \\
				luca.barletta@polimi.it}
			\and
			\IEEEauthorblockN{Elza Erkip}
			\IEEEauthorblockA{\textit{NYU} \\
				New York, USA \\
				elza@nyu.edu}
			\and
			\IEEEauthorblockN{Yonina C. Eldar}
			\IEEEauthorblockA{\textit{	Technion}\\
				Haifa, Israel \\
				yonina@ee.technion.ac.il}

		}
	}
	
	\maketitle

	\begin{abstract}
The capacity of a Multiple-Input Multiple-Output (MIMO) channel in which the antenna outputs are processed by an analog linear combining network and quantized by a set of threshold quantizers is studied. The linear combining weights and quantization thresholds are selected from a set of possible configurations as a function of the channel matrix. The possible configurations of the combining network model specific analog receiver architectures, such as single antenna selection, sign quantization of the antenna outputs or linear processing of the outputs. An interesting connection between the capacity of this channel and a constrained sphere packing problem in which unit spheres are packed in a hyperplane arrangement is shown. From a high-level perspective, this follows from the fact that each threshold quantizer can be viewed as a hyperplane partitioning the transmitter signal space. Accordingly, the output of the set of quantizers corresponds to the possible regions induced by the hyperplane arrangement corresponding to the channel realization and receiver configuration. This connection provides a number of important insights into the design of quantization architectures for MIMO receivers; for instance, it shows that for a given number of quantizers, choosing configurations which induce a larger number of partitions can lead to higher rates.\footnote{This work has been supported in part by NSF Grant \#1527750.}

	\end{abstract}
	\vspace{-0.1cm}
	\section{Introduction}
	As the coupling of multiple antennas and low-resolution quantization hold the promise of enabling millimeter-wave communication, 
		the effect of finite-precision output quantization on the performance of MIMO systems has been widely investigated in recent literature. 
	In \cite{rini2017generalITW}, the authors propose a general framework to study  the capacity of MIMO channels with various output quantization constraints and derive some initial results on the scaling of capacity in the number of available quantization levels.
	In this paper, we further our understanding of output quantization constraints in MIMO channels 
	by drawing a connection between a constrained sphere packing problem and formulation in \cite{rini2017generalITW}.
	This connection suggest a rather insightful geometric-combinatorial approach to the design of receiver quantization strategies for MIMO channel with output quantization.

	\subsubsection*{Literature Review}
	In~\cite{singh2009limits}, low resolution output quantization for MIMO channels is investigated through numerical evaluations. The authors are perhaps the first to note that the loss due to quantization can be relatively small.
	Quantization for the SISO channel is  studied in detail where it is shown that, if the output is quantized using $M$
	bits, then  the optimal input distribution need not have more than $M + 1$ points in its support. 
	A cutting-plane algorithm is employed to compute this capacity and to generate optimum input support. 
	In \cite{mo2015capacity}, the authors study the capacity of  MIMO channels with sign quantization of the outputs and reveal a connection between a geometric-combinatorial problem and the capacity of this model at high SNR.
	\subsubsection*{Contributions}
	In the model of \cite{rini2017generalITW}, the output of a MIMO channel is processed by an analog combining network before being fed to $\Ntq$ threshold quantizers.
	The combining network is chosen among a set of possible configurations as a function of the channel matrix: these configurations represent analog operations that can be performed by the receiver analog front-end. 
	Through the problem formulation in \cite{rini2017generalITW}, it is possible to study the performance of different receiver architectures as a function of the available quantization bits $\Ntq$ and transmit power.

	In this paper, we show that the capacity of the model in \cite{rini2017generalITW} can be approximately characterized using the solution of a geometric-combinatorial problem.
	Each threshold quantizer in effect obtains a linear combination of the noisy channel inputs and can thus be viewed as partitioning the transmit signal space with a hyperplane.
    The output of the set of quantizers identifies a region among those induced  by the hyperplane arrangement corresponding to the channel matrix and receiver configuration.  
	Transmitted points can be reliably distinguished at the receiver when they are separated by a hyperplane in the transmit space. 
	Our result generalizes those of \cite{rini2017generalITW,mo2015capacity} and provides an intuitive approach to design  effective, and sometimes surprising, quantization strategies.
	For example, one would expect that, for a receiver able to perform linear combination before quantization, the optimal transmission strategy is to perform Singular Value Decomposition (SVD)  followed by multilevel quantization of each sub-channel. 	
	We show that this scheme is actually sub-optimal at high SNR as receiver configurations which induce a larger number of partitions may lead to  higher transmission rates.

	\subsubsection*{Organization}
	The channel model is introduced in Sec. \ref{sec:Channel Model}. Combinatorial notions are presented in Sec. \ref{Combinatorial Interlude}. 
	Prior results and a  motivating example given in Sec. \ref{sec:Relevant Architectures}. 
	The main result is presented in Sec. \ref{sec:Main Result}.
	Sec. \ref{sec:Conclusion} concludes the paper. 

	\subsubsection*{Notation}
	All logarithms are taken in base two.
	The vector $\diag\{\Mv\}$ is the diagonal of the matrix $\Mv$ while $\la(\Mv)$  is the vector of eigenvalues of $\Mv$. 
	The  identity matrix  of size $n$ is  $\Iv_{n}$, the  all--zero/all--one matrix of size $n \times m$ is $\zerov_{n \times m}$/$\onev_{n \times m}$, respectively.
Dimensions for these matrices are omitted when implied by the context. We adopt the conve1ntion that  {\small $\fact{n}{i} = 0$ } for $i > n$.
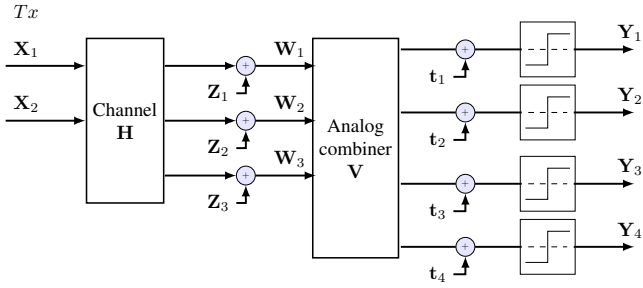
\begin{figure}
	\centering
	\resizebox{9 cm }{!}{
		\label{key}
		\begin{tikzpicture}[node distance=2.5cm,auto,>=latex]
\tikzset{point/.style={coordinate},
	block/.style ={draw, thick, rectangle, minimum height=4em, minimum width=6em},
	line/.style ={draw, very thick,-},
}

\node  at (-4,4) {$Tx$} ;

\node  at (-4.5,3) (x1){} ;
\node  at (-4.5,2) (x2){} ;
\node  at (-4,3.3) {$\Xv_1$} ;
\node  at (-4,2.3) {$\Xv_2$} ;
\node  at (-2.8,3) (hi1){} ;
\node  at (-2.8,2) (hi2){} ;
\draw (x1) [->,line width=1 pt] -> (hi1);
\draw (x2) [->,line width=1 pt] -> (hi2);
\node (b2) [block, minimum height = 3 cm, minimum width=1 cm,thick =2cm, align=center] at(-2.2, 2){\textrm{Channel} \\
${\Hv}$};
\node  at (-1.6,3) (h1){} ;
\node  at (-1.6,2) (h2){} ;
\node  at (-1.6,1) (h3){} ;
\node  at (0.8,3.3) (w1){$\Wv_1$};
\node  at (0.8,2.3) (w2){$\Wv_2$};
\node  at (0.8,1.3) (w3){$\Wv_3$};
\node  at (1.4,3) (w1){} ;
\node  at (1.4,2) (w2){} ;
\node  at (1.4,1) (w3){} ;

\node (b4-align) [block, minimum height = 4 cm, minimum width=1 cm,thick =2cm, align=center] at(2, 1.5){\textrm{Analog}\\ \textrm{combiner}\\   ${{\Vv}}$};

\node  at (0,3) (e1) [sum,scale=0.5]{$+$};
\node  at (0,2) (e2) [sum,scale=0.5]{$+$};
\node  at (0,1) (e3) [sum,scale=0.5]{$+$};

\draw (h1) [->,line width=1 pt] -> (e1);
\draw (h2) [->,line width=1 pt] -> (e2);
\draw (h3) [->,line width=1 pt] -> (e3);
\draw (e1) [->,line width=1 pt] -> (w1);
\draw (e2) [->,line width=1 pt] -> (w2);
\draw (e3) [->,line width=1 pt] -> (w3);
\node  at (7.2,3+0.3) (t1){} ;
\node  at (7.2,2+0.15) (t2){} ;
\node  at (7.2,1-0.15) (t3){} ;
\node  at (7.2,0-0.3) (t4){} ;

\node  at (7,3.3+0.3)  {$\Yv_1$};
\node  at (7,2.3+0.15)  {$\Yv_2$};
\node  at (7,1.3-0.15)  {$\Yv_3$};
\node  at (7,0.3-0.3)  {$\Yv_4$};

\node  at (2.7,3 + 0.3) (y1) {};
\node  at (2.7,2+0.15) (y2) {};
\node  at (2.7,1-0.15) (y3) {};
\node  at (2.7,0-0.3) (y4) {};

\node  at (4,3+0.3) (e11) [sum,scale=0.5]{$+$};
\node  at (4,2+0.15) (e21) [sum,scale=0.5]{$+$};
\node  at (4,1-0.15) (e31) [sum,scale=0.5]{$+$};
\node  at (4,0-0.3) (e41) [sum,scale=0.5]{$+$};

\node  at (5.15,3+0.3) (quant1i){} ;
\node  at (5.85,3+0.3) (quant1o){} ;
\drawrecord(5.5,3+0.3){}

\node  at (5.15,2+0.15) (quant2i){} ;
\node  at (5.85,2+0.15) (quant2o){} ;
\drawrecord(5.5,2+0.15){}

\node  at (5.15,1-0.15) (quant3i){} ;
\node  at (5.85,1-0.15) (quant3o){} ;
\drawrecord(5.5,1-0.15){}

\node  at (5.15,0-0.3) (quant4i){} ;
\node  at (5.85,0-0.3) (quant4o){} ;
\drawrecord(5.5,0-0.3) {}

\draw (y1) [line width=1 pt] -> (e11);
\draw (y2) [line width=1 pt] -> (e21);
\draw (y3) [line width=1 pt] -> (e31);
\draw (y4) [line width=1 pt] -> (e41);
\draw (e11) [line width=1 pt] -> (quant1i);
\draw (e21) [line width=1 pt] -> (quant2i);
\draw (e31) [line width=1 pt] -> (quant3i);
\draw (e41) [line width=1 pt] -> (quant4i);

\draw (quant1o) [->,line width=1 pt] -> (t1);
\draw (quant2o) [->,line width=1 pt] -> (t2);
\draw (quant3o) [->,line width=1 pt] -> (t3);
\draw (quant4o) [->,line width=1 pt] -> (t4);
\node  at (-0.5,3-.5) (n1) {$\Zv_1$};
\node  at (-0.5,2-.5) (n2) {$\Zv_2$};
\node  at (-0.5,1-.5) (n3) {$\Zv_3$};
\draw (n1) [->,line width=1 pt] -| (e1) ;
\draw (n2) [->,line width=1 pt] -| (e2);
\draw (n3) [->,line width=1 pt] -| (e3);

\node  at (4-0.5,3-.5+0.3) (th1) {$\tv_1$};
\node  at (4-0.5,2-.5+0.15) (th2) {$\tv_2$};
\node  at (4-0.5,1-.5-0.15) (th3) {$\tv_3$};
\node  at (4-0.5,0-.5-0.3) (th4) {$\tv_4$};
\draw (th1) [->,line width=1 pt] -| (e11);
\draw (th2) [->,line width=1 pt] -| (e21);
\draw (th3) [->,line width=1 pt] -| (e31);
\draw (th4) [->,line width=1 pt] -| (e41);

\end{tikzpicture}	%
	}
	\caption{System model for $\Ntx = 2$, $\Nrx = 3$, and $\Ntq = 4$.}
	\label{fig:system mOdel}
	\vspace{-0.5 cm}
\end{figure}

\section{Channel Model}
\label{sec:Channel Model}
Consider the discrete-time MIMO channel with $\Ntx$/$\Nrx$ transmit/receive antennas in which an input
vector $\Xv_n=[X_{1,n} \ldots X_{\Ntx,n}]^T$ results in the output vector $\Wv_n=[W_{1,n} \ldots W_{\Nrx,n}]^T$ 
according to the relationship
\ea{
	\Wv_n=\Hv \Xv_n+\Zv_n, \quad  1 \leq n \leq N,
	\label{eq:antenna out}
}
where  $\Zv_n$ is an  i.i.d.  Gaussian vector of size $\Nrx$ with zero mean and unit variance
and
 $\Hv$ is a full rank matrix of size $\Nrx \times \Ntx$ (i.e. $\rank(\Hv) = \min(\Ntx, \Nrx)$), fixed
through the transmission block-length and  known at the receiver and transmitter.\footnote{The full rank assumption is justified for richly scattering environments.} The input is subject to the power constraint $\sum_{n=1}^N \Ebb[ |\Xv_{n}|_2^2]\leq  N P$ where $|\Xv_{n}|_2$ is the 2-norm. 
We study a variation of the model in \eqref{eq:antenna out} shown in Fig.\ref{fig:system mOdel} in which the output vector $\Wv_n$ is  processed by a receiver analog front-end  and 
fed to $\Ntq$ threshold one-bit quantizers. This results in 
the vector  $\Yv_n= [Y_{1,n} \ldots Y_{\Ntq,n}]^T \in \{-1,+1\}^{\Ntq}$ given by
\ea{
\Yv_n = \sign(\Vv \Wv_n+\tv), \quad   1 \leq n \leq N,
\label{eq:quantized model}
}
where $\Vv$ is the analog combining matrix of size $\Ntq \times \Nrx$, $\tv$~is a threshold vector of length $\Ntq$ and
$\sign(\uv)$ is the function producing the sign of each component of the vector $\uv$ as plus or minus one.
The matrix $\Vv$ and the vector $\tv$ are selected among a set of available configurations $\Fcal$ \cite{rini2017generalITW}:
\ea{
	\lcb \Vv,\tv \rcb \in \Fcal \subseteq \lcb \Rbb^{\Ntq \times \Nrx} ,	 \Rbb^{\Ntq} \rcb.
	\label{eq:available conf}
}
For a fixed receiver configuration, $\{\Vv, \tv\}$, the capacity of the channel in \eqref{eq:quantized model} is obtained as
\ea{
\Ccal(\Vv,\tv)=\max_{P_{\Xv}(\xv), \ \Ebb[ |\Xv|_2^2]\leq  P} I(\Xv;\Yv).
\label{eq:capacity V t}
}
We are interested in determining the largest attainable performance over all possible configurations, leading to
\ea{
\Ccal(\Fcal)=\max_{\{\Vv, \tv\} \in \Fcal} \Ccal(\Vv,\tv).
\label{eq:capacity F}
}

In the following, we provide an approximate characterization of the solution of the maximization in \eqref{eq:capacity F} under the assumption that 
$\diag\{\Hv \Hv^T\}=\diag\{\Vv \Vv^T\}=\ones_{1 \times \Nrx}.$
Under this assumption, the derivation of the results is particularly succinct and thus fitting to the available space.
The more general case of any arbitrary channel matrix $\Hv$ and any combining matrix $\Vv$ is presented in \cite{UStranDS}.

\section{Combinatorial Interlude}
\label{Combinatorial Interlude}
This section briefly introduces a few combinatorial concepts useful in the remainder of the paper. 

A  \emph{hyperplane arrangement}  $\Acal$ is a finite set of $n$ affine hyperplanes in $\Rbb^m$ for some $n,m \in \Nbb$.
A hyperplane arrangement  $\Acal=\{ \xv \in \Rbb^m, \ \av_i^T \xv = b_i \}_{i=1}^n$ can be expressed as $\Acal=\lcb \xv,  \  \Av \xv=\bv\rcb$
where $\Av$ is obtained by letting each row $i$ correspond to $\av_i^T$ and defining $\bv=[b_1 \ldots b_n]^T$.
A  plane arrangement is said to be in General Position (GP) if and only if every $n \times n$ sub-matrix of  $\Av$ has non zero determinant \cite{Tcover1965Geometrical}.

\begin{lem}
	\label{lem:hyperplane}
	A hyperplane arrangement of size $n$  in  $\Rbb^m$ divides $\Rbb^m$ into at most 
	{\small $\rr(m,n)=\sum_{i=0}^{m} \fact{n}{i} \leq 2^n$}
	regions.
\end{lem}
\begin{lem}
	\label{lem:origin arrangment}
	A hyperplane arrangement of size $n$  in  $\Rbb^m$  where all the hyperplanes pass through the origin 
	divides $\Rbb^m$ into at most  {\small $\ror(n,m)= 2\sum_{i=0}^{m	} {{n-1}\choose{i}}$ }
regions.
\end{lem}

\begin{lem}
	\label{lem:pcomp}
	Let $\Acal$ be a hyperplane arrangement of size  $l$ in $\Rbb^m$ and consider a hyperplane arrangement $\Bcal$ of size $dl$ with $d \in \Nbb$ hyperplanes parallel to each of the hyperplanes in $\Acal$. Then $\Bcal$ divides $\Rbb^m$ into at most	
	$\rpar(m,n,d)= \sum_{i=0}^{m} \binom{l}{i} d^i  \leq (1+d)^l$
	regions.
\end{lem}
A necessary condition to attain the qualities in Lem. \ref{lem:hyperplane},  Lem. \ref{lem:origin arrangment}  and Lem. \ref{lem:pcomp}
is for the hyperplane arrangement $\Acal$ to be in GP. 
A unitary sphere packing in $\Rbb^m$ is defined as 
\ea{
	\Pcal=\bigcup_{i=1}^N \Scal^m(c_i,1),
}
where $\Scal^m(c,r)$ is the $m$-dimensional hyper-sphere with center $c$ and radius $r$. 
A hyperplane separates two spheres if each sphere belongs to one of the half-spaces induced by the hyperplane.
A sphere packing $\Pcal$  is said to be separable by the hyperplane arrangement $\Acal$ if  any two spheres are separated by at least one hyperplane in $\Acal$.  
A sphere packing in a sphere  is a packing $\Pcal$ for which $\Pcal \subseteq \Scal(c,r)$ for some $c,r$. 

Our aim is to show a connection between the capacity in \eqref{eq:capacity F} and the following sphere packing problem. 

\begin{definition}{\textit{ Separable sphere packing in a sphere:}}
	\label{def:Separable sphere packing in a sphere}
	Given a hyperplane arrangement $\Acal$ and a constant $r \in \Rbb^+$, define $\rssps(\Acal,r)$ as the largest number of unit spheres in a packing $\Pcal$ contained in the sphere $\Scal(\zerov,r)$ and separable by $\Acal$.
\end{definition}

\section{Prior Results and a Motivating Example}
\label{sec:Relevant Architectures}
The maximization in \eqref{eq:capacity F} yields the optimal performance for any set of possible receiver configurations. 
One is often interested in studying and comparing the performance for specific classes of receiver configurations: three such classes are studied in detail in
\cite{rini2017generalITW}:
single antenna selection and multilevel quantization, sign quantization of the outputs and  linear combining and multilevel quantization.

\subsection{Prior Results}
\label{sec:Prior Results}
The simplest receiver architecture of interest is perhaps the one in which a single antenna output is selected by the receiver and quantized through a high-resolution quantizer. 
This is obtained in the model of  Sec. \ref{sec:Channel Model} by setting
{
\ea{
\Fcal & = \lcb 
\Vv= \lsb \zeros_{\Ntq \times i}~ \ones_{\Ntq \times 1} ~ \zeros_{\Ntq \times\Nrx-i-1} \rsb, \ {\tiny 0 \leq i \leq \Nrx-1 }
\rnone  \nonumber
\\
& \quad \quad \lnone \quad \tv \in \Rbb^{\Ntq} \rcb,
\label{eq:fcal a}
}
}
where the term $\onev_{N_{t_q}\times 1}$ selects the antenna with the highest channel gain.

\begin{prop}{\bf \cite[Prop. 2]{rini2017generalITW}.}
	\label{prop:mimo single}
	The capacity of the MIMO channel with single antenna selection and multilevel quantization is upper--bounded as
	\ea{ \label{eq:mimo_b}
		\Ccal_{\rm select}\leq \f 12 \log \lb \min \lcb 1+|\hv_{\max}^T|_2^2P,(\Ntq+1)^2 \rcb \rb,
	}
	where $\hv_{\max}^T$ is the row of $\Hv$ with the largest 2-norm. The upper bound in~\eqref{eq:mimo_b}
	can be attained to within $2$ bits-per-channel-use ($\bpcu)$. 
\end{prop}

Our main result, discussed in detail in Sec.\ref{sec:Main Result}, is inspired by an intriguing connection between Lem. \ref{lem:origin arrangment} and the infinite SNR capacity of the MIMO channel
with  sign quantization of the  outputs \cite{mo2015capacity}.
Note that the model in \cite{mo2015capacity} is obtained by setting $\Ntq=\Nrx$ and letting $\Fcal$ be the set of all matrices obtained by permuting the rows of $[\Iv, \zerov]$.
\begin{prop}{\bf \cite[Prop. 3]{mo2015capacity}.}
	\label{prop:finite capacity}
	The capacity of the  MIMO  channel with sign quantization of the outputs in which $\Hv$ is in GP
	at infinite SNR is bounded as
	\ean{
	\log(	\ror(\Nrx,\Ntx) ) \leq \Ccal^{\rm SNR \goes \infty}_{\sign} \leq  \log(\ror(\Nrx,\Ntx)+1).
	}
\end{prop}

Recall that the most general architecture in Sec. \ref{sec:Channel Model} has
\ea{
\Fcal= \lcb \Vv \in \Rbb^{\Ntq \times \Nrx}, \ \tv \in \Rbb^{\Ntq} \rcb,
\label{eq:fcal b}
}
and corresponds to a receiver analog front-end which is able to perform any linear combination of the antenna outputs before quantization. 
\begin{prop}{\bf \cite[Prop. 6]{rini2017generalITW}. }
	\label{prop:mimo linear}
	The capacity of a MIMO channel with linear combining and multilevel quantization is upper--bounded as
	\ea{
		& \Ccal_{\rm linear} \leq  R^\star(\la(\Hv),P,\Ntq)+K.
		\label{eq:mimo linear}
	}
	The capacity is to within a gap of  $3 K \ \bpcu$ from the upper bound in \eqref{eq:mimo linear} for 
	\small \ea{
		& R^\star(\la(\Hv),P,\Ntq)= \nonumber \\
		& \lcb \p{
			\sum_{i=1}^{K}\f 12 \log (1+\la_i P_i)  \\
			\quad \quad  \quad   \quad \quad  \quad \quad  {\rm{if}} \ \  \sum_{i=1}^{K} \lb \sqrt{1+\la_i^2 P_i}-1 \rb \leq \Ntq \\
			K  \log \lb \f {\Ntq} K  +1 \rb\\
			\quad \quad  \quad  \quad \quad  \quad \quad   {\rm \small otherwise},
		}\rnone
		\label{eq:general waterfilling}
	}\normalsize
with $K=\max \{N_t,N_r\}$, $P_i=(\mu-\la_i^{-2})^+$ and $\mu \in \Rbb^+$ is the smallest value for which  $\sum_i P_i=P$.
\end{prop}
To establish the achievability of Prop. \ref{prop:mimo linear}, the SVD can be used to transform the channel into  $K=\min \{N_t,N_r\}$ parallel sub-channels with independent unit-variance additive noise and gains $\la(\Hv)$.
After SVD, the quantization strategy is chosen depending on whether the performance is bounded by the effect of the additive noise or by the quantization noise. 
\subsection{Motivating Example for the Combinatorial Approach}
\label{sec:Motivating Example}
Let us consider the three architectures in Propositions 1-3 
for the case of $\Ntx=2$, $\Nrx=3$ and $\Ntq=4$,  also shown in Fig. \ref{fig:system mOdel}, and provide some high-level intuition on the relationship between capacity and the sphere packing problem in Def. \ref{def:Separable sphere packing in a sphere}.

{\bf Prop. \ref{prop:mimo single}:} Since the threshold quantizers are used to sample the same antenna output, the number of possible outputs is at most $\Nrx+1$ so that the performance in  Prop. \ref{prop:mimo single} is bounded by $\log(\Ntq+1)=\log 5 \approx 2.32 \ \bpcu$ at high SNR.
This receiver configuration can be interpreted as follow:
an antenna output represents a line in the two-dimensional transmit signal space; each threshold quantizer  corresponds to a translation of this line and these $\Ntq$ parallel lines partition the signal space into at most $\Ntq+1$ subregions.

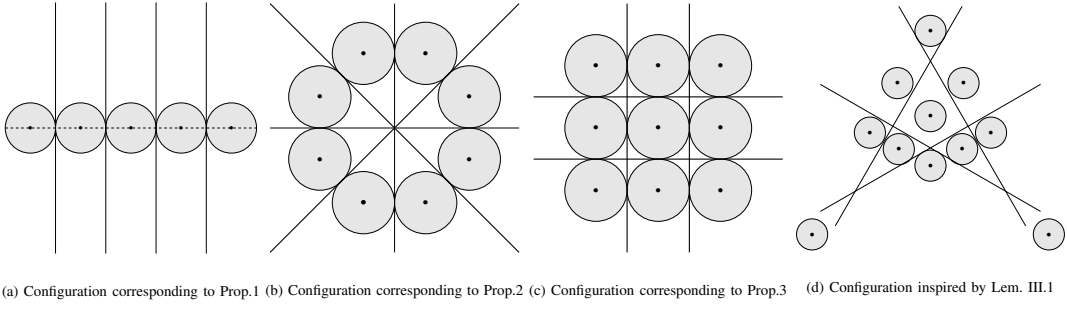
\begin{figure*}
\vspace*{0cm}
\captionsetup[subfigure]{justification=centering}
\centering
\begin{adjustbox}{scale = 0.63}

\begin{subfigure}[d]{0.3 \textwidth}
\resizebox{\textwidth}{!}{

\begin{tikzpicture}[node distance=2.5cm,auto,>=latex]
		\tikzset{point/.style={coordinate},
			block/.style ={draw, thick, rectangle, minimum height=4em, minimum width=6em},
			line/.style ={draw, very thick,-},
		}
		\draw[dashed] (-5,0) [line width=1/4 pt ] -> (5,0);
		
		\draw (1,-5) [line width=1 pt ] -> (1,5);

		\draw (-1,-5) [line width=1 pt ] -> (-1,5);

		\draw (-3,-5) [line width=1 pt ] -> (-3,5);

		\draw (3,-5) [line width=1 pt ] -> (3,5);

		\draw[solid, draw=black, ] (-4,0) circle (1);
		\fill[solid, draw=black,opacity=0.1] (-4,0) circle (1);

		\draw[solid, draw=black, ] (4,0) circle (1);
		\fill[solid, draw=black,opacity=0.1] (4,0) circle (1);

		\draw[solid, draw=black, ] (-2,0) circle (1);
		\fill[solid, draw=black,opacity=0.1] (-2,0) circle (1);

		\draw[solid, draw=black, ] (2,0) circle (1);
		\fill[solid, draw=black,opacity=0.1] (2,0) circle (1);

		\draw[solid, draw=black, ] (0,0) circle (1);
		\fill[solid, draw=black,opacity=0.1] (0,0) circle (1);

        \foreach \Point in {(-4,0), (4,0), (-2,0), (2,0),(0,0)}{
        \node at \Point {\textbullet};
}
		\end{tikzpicture}

  	}
	\vspace{0cm}
	\caption{Configuration corresponding to Prop.\ref{prop:mimo single}}
	\label{fig:a}
\end{subfigure}

\begin{subfigure}[d]{0.3 \textwidth}
	\resizebox{\textwidth }{!}{
		\begin{tikzpicture}[node distance=2.5cm,auto,>=latex]
		\tikzset{point/.style={coordinate},
			block/.style ={draw, thick, rectangle, minimum height=4em, minimum width=6em},
			line/.style ={draw, very thick,-},
		}
		\node  at (0,0) (cent5er) {};
		\draw (-4,0) [line width=1/4 pt ] -> (4,0);
		
		\draw (0,-4) [line width=1/4 pt] -> (0,4);
		
		\draw (-4,-4) [line width=1 pt ] -> (4,4);

		\draw (4,-4) [line width=1 pt ] -> (-4,4);

		\draw[solid, draw=black, ] (2.4,1) circle (1);
		\fill[solid, draw=black,opacity=0.1] (2.4,1) circle (1);
		
		\draw[solid, draw=black, ] (-2.4,1) circle (1);
		\fill[solid, draw=black,opacity=0.1] (-2.4,1) circle (1);
		
		\draw[solid, draw=black, ] (1,2.4) circle (1);
		\fill[solid, draw=black,opacity=0.1] (1,2.4) circle (1);
		
		\draw[solid, draw=black, ] (-1,2.4) circle (1);
		\fill[solid, draw=black,opacity=0.1] (-1,2.4) circle (1);
		
		\draw[solid, draw=black, ] (-2.4,-1) circle (1);
		\fill[solid, draw=black,opacity=0.1] (-2.4,-1) circle (1);
		
		\draw[solid, draw=black, ] (2.4,-1) circle (1);
		\fill[solid, draw=black,opacity=0.1] (2.4,-1) circle (1);
		
		\draw[solid, draw=black, ] (-1,-2.4) circle (1);
		\fill[solid, draw=black,opacity=0.1] (-1,-2.4) circle (1);
		
		\draw[solid, draw=black, ] (1,-2.4) circle (1);
		\fill[solid, draw=black,opacity=0.1] (1,-2.4) circle (1);

		\foreach \Point in {(1,-2.4), (-1,-2.4), (2.4,-1), (-2.4,-1), (-1,-2.4),(-1,2.4), (1,2.4),(-2.4,1),(2.4,1)}{
        \node at \Point {\textbullet};
}		
		
		\end{tikzpicture}

  	}
	\vspace{0cm}
	\caption{Configuration corresponding to Prop.\ref{prop:finite capacity}}
	\label{fig:origin}
\end{subfigure}

 \begin{subfigure}[d]{0.3 \textwidth}
        \resizebox{\textwidth }{!}{
		\begin{tikzpicture}[node distance=2.5cm,auto,>=latex]
		\tikzset{point/.style={coordinate},
			block/.style ={draw, thick, rectangle, minimum height=4em, minimum width=6em},
			line/.style ={draw, very thick,-},
		}
		\draw (-1,-4) [line width=1 pt ] -> (-1,4);

		\draw (1,-4) [line width=1 pt ] -> (1,4);
		
		\draw (-4,-1) [line width=1 pt ] -> (4,-1);
				
		\draw (-4,1) [line width=1 pt ] -> (4,1);
		
		\draw[solid, draw=black, ] (0,0) circle (1);
		\fill[solid, draw=black,opacity=0.1] (0,0) circle (1);
		
		\draw[solid, draw=black, ] (2,0) circle (1);
		\fill[solid, draw=black,opacity=0.1] (2,0) circle (1);

    	\draw[solid, draw=black, ] (-2,0) circle (1);
		\fill[solid, draw=black,opacity=0.1] (-2,0) circle (1);

    	\draw[solid, draw=black, ] (0,2) circle (1);
		\fill[solid, draw=black,opacity=0.1] (0,2) circle (1);

    	\draw[solid, draw=black, ] (0,-2) circle (1);
		\fill[solid, draw=black,opacity=0.1] (0,-2) circle (1);

        \draw[solid, draw=black, ] (2,-2) circle (1);
		\fill[solid, draw=black,opacity=0.1] (2,-2) circle (1);
		
		\draw[solid, draw=black, ] (2,2) circle (1);
		\fill[solid, draw=black,opacity=0.1] (2,2) circle (1);
		
		\draw[solid, draw=black, ] (-2,-2) circle (1);
		\fill[solid, draw=black,opacity=0.1] (-2,-2) circle (1);

		\draw[solid, draw=black, ] (-2,2) circle (1);
		\fill[solid, draw=black,opacity=0.1] (-2,2) circle (1);
    
    		\foreach \Point in {(0,0), (2,0), (-2,0), (0,2), (0,-2),(2,-2), (2,2),(-2,-2),(-2,2)}{
        \node at \Point {\textbullet};
        }

		\end{tikzpicture}
	}
\vspace{0cm}

\caption{Configuration corresponding to Prop.\ref{prop:mimo linear}}
\label{fig:channel model sign}
 \end{subfigure}
\begin{subfigure}[d]{0.32 \textwidth}
	\resizebox{\textwidth}{!}{
		\begin{tikzpicture}[node distance=2.5cm,auto,>=latex]
		\tikzset{point/.style={coordinate},
			block/.style ={draw, thick, rectangle, minimum height=4em, minimum width=6em},
			line/.style ={draw, very thick,-},
		}
		\node  at (0,0) (center) {};
		\draw (-4,-3.464) [line width=1 pt ] -> (4,1.155);

		\draw (-4,1.155) [line width=1 pt ] -> (4,-3.464);
		
		\draw (-3.46,-3.993) [line width=1 pt ] -> (1.15,3.992);
		
		\draw (3.46,-3.993) [line width=1 pt ] -> (-1.15,3.992);
		
		\draw[solid, draw=black, ] (0,0) circle (0.565);
		\fill[solid, draw=black,opacity=0.1] (0,0) circle (0.565);
		
		\draw[solid, draw=black, ] (0,3.1) circle (0.565);
		\fill[solid, draw=black,opacity=0.1] (0,3.1) circle (0.565);
		
		\draw[solid, draw=black, ] (0,-1.8) circle (0.565);
		\fill[solid, draw=black,opacity=0.1] (0,-1.8) circle (0.565);
		
		\draw[solid, draw=black, ] (2.2,-0.6) circle (0.565);
		\fill[solid, draw=black,opacity=0.1] (2.2,-0.6) circle (0.565);
		
		\draw[solid, draw=black, ] (-2.2,-0.6) circle (0.565);
		\fill[solid, draw=black,opacity=0.1] (-2.2,-0.6) circle (0.565);
		
		\draw[solid, draw=black, ] (1.15,-1.2) circle (0.565);
		\fill[solid, draw=black,opacity=0.1] (1.15,-1.2) circle (0.565);
		
		\draw[solid, draw=black, ] (-1.15,-1.2) circle (0.565);
		\fill[solid, draw=black,opacity=0.1] (-1.15,-1.2) circle (0.565);
		
		\draw[solid, draw=black, ] (1.2,1.2) circle (0.565);
		\fill[solid, draw=black,opacity=0.1] (1.2,1.2) circle (0.565);
		
		\draw[solid, draw=black, ] (-1.2,1.2) circle (0.565);
		\fill[solid, draw=black,opacity=0.1] (-1.2,1.2) circle (0.565);
		
		\draw[solid, draw=black, ] (-4.3,-4.3) circle (0.565);
		\fill[solid, draw=black,opacity=0.1] (-4.3,-4.3) circle (0.565);
		
		\draw[solid, draw=black, ] (4.3,-4.3) circle (0.565);
		\fill[solid, draw=black,opacity=0.1] (4.3,-4.3) circle (0.565);

		\foreach \Point in {(0,0), (0,3.1), (0,-1.8), (2.2,-0.6), (-2.2,-0.6),(1.15,-1.2),(-1.15,-1.2), (1.2,1.2),(-1.2,1.2),(-4.3,-4.3) ,(4.3,-4.3)}{
        \node at \Point {\textbullet};
        }
		\end{tikzpicture}
	}
	\hspace{1cm}
	\caption{Configuration inspired by Lem. \ref{lem:hyperplane}}
	\label{fig:single antenna}
\end{subfigure}

\end{adjustbox}
\caption{
Different receiver output quantization strategies.
}
\label{fig:receiver models}
\vspace{-.5 cm }
\end{figure*}

{\bf Prop. \ref{prop:finite capacity}:}
Sign quantization of the outputs corresponds to the hyperplane arrangement in which all hyperplanes pass
through the origin: the  number of regions induced by this arrangement is obtained through Lem. \ref{lem:origin arrangment}. There are $\ror=8$ partitions, as also shown in Fig. \ref{fig:origin}, yielding a maximum rate of $3 \ \bpcu$, attainable at high SNR.

{\bf Prop. \ref{prop:mimo linear}:}
When the receiver can perform linear combining before quantization, the SVD can be used to transform the channel into two parallel sub-channels. 
This strategy corresponds to the hyperplane arrangement in Lem. \ref{lem:pcomp} and  the number of partitions induced is $9$, as also shown in Fig. \ref{fig:channel model sign}. 

{\bf Lem. \ref{lem:hyperplane}:}
This lemma actually indicates that the largest number of regions is $11$ so that the rate $\log(11)=3.46 \ \bpcu$ can be obtained through the receiver configuration in Fig. \ref{fig:single antenna} at high SNR.\footnote{Note that this does not contradict the result of Prop. \ref{prop:mimo linear} since the inner bound is $2 \ \bpcu$ from the outer bound.}

Given the above interpretation of the capacity at high SNR,
a feasible finite SNR strategy is the one in which, for a given
receiver configuration, the channel inputs are chosen as the
center of the spheres with sufficiently large radius inside each partition subject to the power constraint. The average achievable rate of the four strategies discussed above is plotted in Fig. 3.
Each line in Fig. 3 corresponds to one of the sphere packing configurations in Fig. 2. For a given channel realization, $\Vv$ and $\tv$ are chosen to result in the partitionings of the transmitter space corresponding to each of the subfigures in Fig. 2, appropriately scaled by the available transmit power. Note that the configurations are not optimized. The channel inputs are then chosen as uniformly distributed over the center of the spheres packed in the partitionings. The average performance is calculated over
real i.i.d. zero-mean, unitary variance Gaussian channel gains,
further scaled to guarantee that each row has unitary 2-norm. The capacity of the channel without quantization constraint is also provided as a reference.

\begin{figure}
	\hspace*{-0.5cm}                                        
	\begin{tikzpicture}[node distance=2.5cm,auto,>=latex] \node at (-2.7,7)
	{\includegraphics[trim={2.5cm 0cm 0 0},scale = 0.27]{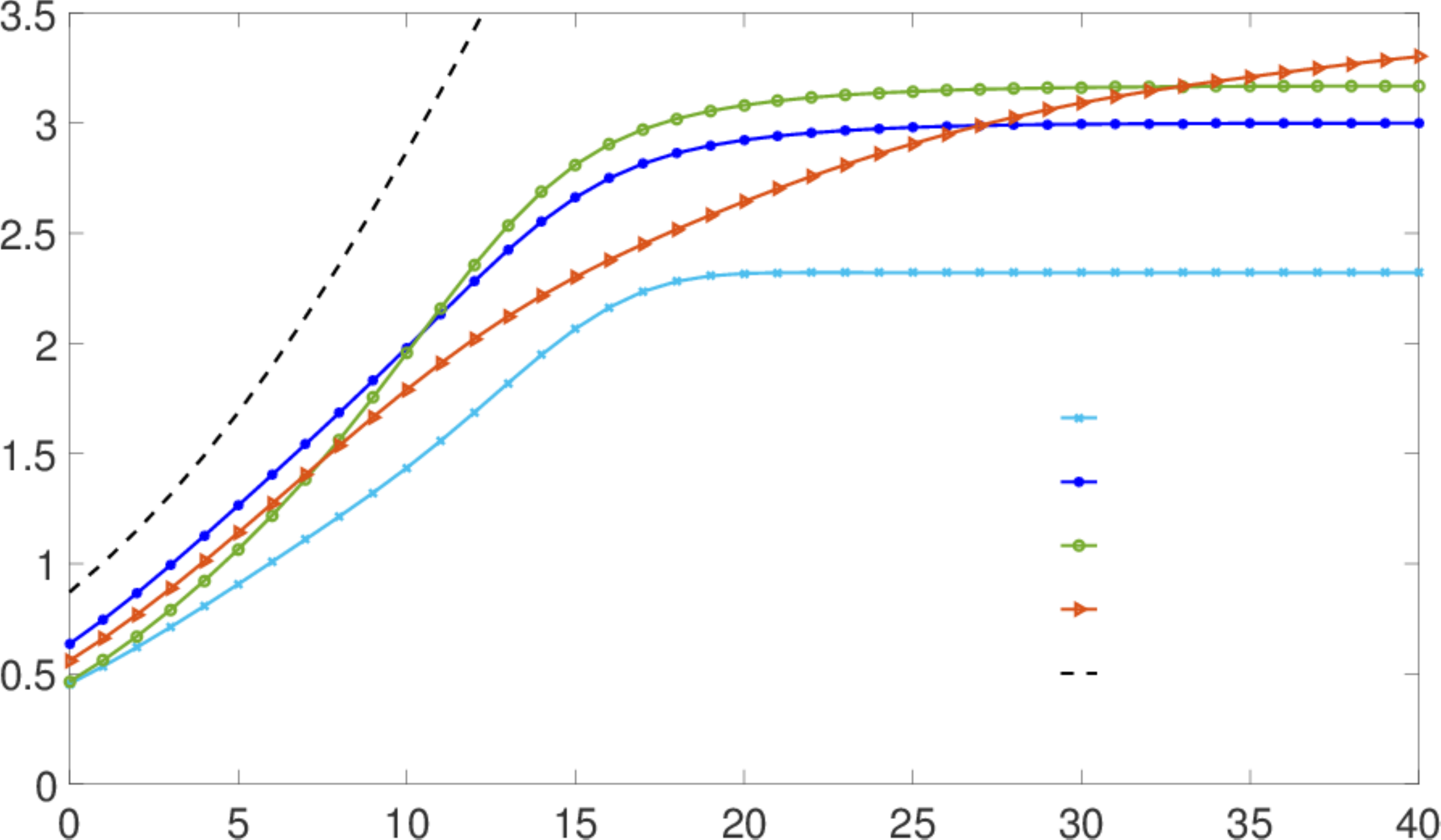}};
	\node at (-7.6,7)  [ rotate=90] {\small{Rate $(\bpcu)$}};
	\node at (-2.5,4.35) {\small{$P$ (dB)}};
	\node at (-0.2,7) {\footnotesize{Fig. \ref{fig:a}}};
	\node at (-0.2,6.6) {\footnotesize{Fig. \ref{fig:origin}}};
	\node at (-0.2,6.2) {\footnotesize{Fig. \ref{fig:channel model sign}}};
	\node at (-0.2,5.85) {\footnotesize{Fig. \ref{fig:single antenna}}};
	\node at (0.03,5.5) {\footnotesize{Outer bound}};
	
	\end{tikzpicture}
	\caption{Simulation results for $\Ntq = 4$, $\Nrx = 3$, and $\Ntx =2$ discussed in Sec. \ref{sec:Motivating Example}.}
	\label{fig:results}
	\vspace{-.5 cm}
\end{figure}

From Fig. \ref{fig:results} we see that, at high SNR, the best performance is attained by the configuration corresponding to Lem. \ref{lem:hyperplane}, since at high SNR the performance is determined by the number of transmitted points. As the SNR decreases, configurations with less transmitted points perform better.

\section{Main Result}
\label{sec:Main Result}
Sec. \ref{sec:Motivating Example}  provides a geometric-combinatorial interpretation of the capacity of the model in \eqref{eq:antenna out}-\eqref{eq:quantized model} for the receiver architectures considered in \cite{rini2017generalITW}.
The main result of the paper is to make such interpretation more rigorous and more general.
\begin{theorem}
\label{thm:main}
The capacity expression in \eqref{eq:capacity F} when $\diag\{\Hv \Hv^T\}=\diag\{\Vv \Vv^T\}=\ones_{1 \times \Nrx} $  is upper bounded as
\ea{
\Ccal(\Fcal) \leq \max_{\Acal} \ \  \log \rssps(\Acal,\sqrt{P})+ \f 3 2 K +3,
\label{eq:approx capacity}
}
for 
\ea{
\Acal \in \{ \xv, \ \Vv \Hv \xv=\tv, \quad (\Vv,\tv) \in \Fcal\},
\label{eq:Acal}
}
and $K=\max\{\Ntx,\Nrx\}$. The capacity is within $2.5 \Ntx \ \bpcu$ from the outer bound in \eqref{eq:approx capacity}.
\end{theorem}

\begin{IEEEproof}
Only the converse proof is presented here while the achievability proof is provided in \cite{UStranDS}. 
Let us choose the input and output alphabets as 
$\Xcal=\Ycal=\lsb 0 : \rr(\Ntx,\Ntq) \rsb$	
and let the channel transition probability be determined by the channel input support and the receiver analog configuration. 
Also, let us define $\sign^*(x)$ as
\ean{
\sign^*(x)=
\begin{cases}
x    &  |x|<1\\
\sign(x) & |x|\geq 1,
\end{cases}
}
and the set $\overline{\Nbb}^m$ as
\ea{
\overline{\Nbb}^m =\lcb  \nvb_{\nv}, \ \ \nvb_{\nv} \in \Scal^m(\nv,1), \  \forall \ \nv \  \in \Nbb^m \rcb,
\label{eq:nbb}
}
that is $\overline{\Nbb}^m$ is composed of a set of points selected from the unit sphere around the integer points in $\Nbb^m$.
Finally, let $Q_{\overline{\Nbb}^m}(\xv)$ be the mapping which assigns each point in $\Rbb^m$ to the closest point in $\overline{\Nbb}^m$ and 
\ean{
& \Wvo^N = \Hv Q_{\overline{\Nbb}^\Ntx}(\Xv^N),  \quad
\Yvo^N =\sign^* \lb \Vv \Wvo^N +\tv \rb,\\ 
& \Ev^N  =\Wv^N-\Wvo^N.
}

Using Fano's inequality, we write
\eas{
	N(R-\ep_N) 
	& \leq I(\Yvo^N,\Ev^N;\Xv^N) 
	\label{eq:error}\\
	& \leq I(\Yvo^N;\Xv^N)+H(\Ev^N)-H(\Zv^N) \nonumber \\
	& = I(\Yvo^N;\Xv^N)+  \f {N \Nrx} 2 \log \f 3 2,
		\label{eq:error 2}
}{\label{eq:error tot}}
where, in \eqref{eq:error}, we have used the fact that we can reconstruct $\Yv^N$ from $\Yvo^N$ and the value of $\Ev^N$.  
In \eqref{eq:error 2}, we used the fact that since $\diag\{\Hv \Hv^T\} = 1$, components of  $\Hv(\Xv- Q_{\overline{\Ntx}^m}(\Xv))$ have support at most $[-1,+1]$. The largest variance of a random variable with finite support is for the case in which the probability distribution is evenly distributed at the end points, so that 
$\var[\Ev_i] \leq 3/2$.  Using the ``Gaussian maximizes entropy'' property, we obtain $H(\Ev_i) \leq 1/2 \log( \pi e 3)$.

From a high-level perspective, \eqref{eq:error tot} shows that the capacity of the channel in \eqref{eq:antenna out}-\eqref{eq:quantized model} is close to the capacity of the channel with no additive noise but in which the input is mapped to $\overline{\Nbb}^m$. 
Next, we show that restricting the input to a peak power constraint, instead of an average power constraint, has a bounded effect on the capacity. 

Let us represent $\Xv_i$ in hyper-geometric coordinates as $\Xv_i=\phi_i |\Xv_i|_2$ for $\phi_i  \in \Scal^{\Ntx}(0,1)$ and $|\phi_i|_2=1$ and define $\Xvh^N$ as 
\ea{
\Xvh_i=\phi_i \lb  |\Xv_i|_2 \mod \lceil \sqrt{P} \rceil  \rb, \  1 \leq i \leq N
}
where $\mod(x)$ indicates the modulus operation; in other words, $\Xvh_i$ has the same direction as $\Xv_i$  but its modulus is folded over $\lceil \sqrt{P} \rceil $. 
Accordingly, define
\ean{
\Wvh^N =\Hv Q_{\overline{\Nbb}^\Ntx}(\Xvh ^N), ~ \Yvh^N =\sign^* \lb \Vv \Wvh^N +\tv \rb,  
}
and use these definitions to further bound the term  $I(\Yvo^N;\Xv^N)$  in \eqref{eq:error 2} as
\ea{
	I(\Yvo^N;\Xv^N)   
	& \leq I(\Yvh^N,\Yvo^N;\Xvh^N,\Xv^N)  \label{eq:p22} \\
	& = I(\Yvh^N; \Xvh^N)+I(\Yvh^N;\Xv^N|\Xvh^N)\\
	&+I(\Yvo^N;\Xv^N, \Xvh^N|\Yvh^N).\nonumber
}
Note that $I(\Yvh^N;\Xv^N|\Xvh^N) = 0$ because of the Markov chain $\Yvh^N-\Xvh^N-\Xv^N$. 
For the term $I(\Yvo^N;\Xv^N, \Xvh^N|\Yvh^N) $ we write
\eas{
	I(\Yvo^N;\Xv^N, \Xvh^N|\Yvh^N) 
	& \leq  H(\Yvo^N-\Yvh^N)
	\label{eq:p1}\\
	& \leq H(\Hv (\Xvo^N-\Xvh^N))
	\label{eq:p2} \\
	& \leq H(\Xvo^N-\Xvh^N),
		\label{eq:p3} 
}
where \eqref{eq:p1} follows from the fact that $\Yvo^N$ is a discrete random variable, \eqref{eq:p2} from the fact that $\Xvo$ and $\Xvh$ are also discrete random variables, and \eqref{eq:p3} from the fact that $\Hv$ is full rank by assumption. 

Next, to bound  the term $H(\Xvo^N-\Xvh^N)$, we can write  
\ea{
\Xvo_i-\Xvh_i = \phi_i  \lb |\Xvo_i|_2 \ {\rm /} \ \lceil \sqrt{P} \rceil \rb,
} 
where ${\rm /}$ indicates the quotient of the modulus operation. 
The entropy of this random variable can then be rewritten as 
\ean{
H(\Xvo^N-\Xvh^N) & \leq H(\phi^N)+H(|\Xvo^N|_2 \ {\rm /} \ \lceil \sqrt{P} \rceil ) \\
& \leq N(\Nrx-1) +N \max_i H(|\Xvo_i|_2 \ {\rm /} \ \lceil \sqrt{P} \rceil ).
}
It can be shown that $H(|\Xvo_i|_2 \ {\rm /} \ \lceil \sqrt{P} \rceil ) \leq 3 \  \bpcu$: the proof follows from the fact that the power constraint can be violated only a finite number of times, which leads to the fact that $|\Xvo_i|_2 \ {\rm /} \ \lceil \sqrt{P} \rceil$ is concentrated around small integer values. 
By combining the bounds in \eqref{eq:error tot} and \eqref{eq:p22} we obtain
\ea{
N(R-\ep_N) & \leq I(\Yvh^N; \Xvh^N)+\f 3 2 N K +3 N \nonumber \\
& \leq N \max_{P_{\Xvh}} I(\Yvh; \Xvh)+\f 3 2 N K +3 N.
\label{eq:tot outer}
} 

Let us now evaluate the mutual information term $I(\Yvh^N; \Xvh^N)$, $\Yvh^N$  is a deterministic function of $\Xvh^N$ and can be interpreted as the membership function indicating to which partition of the hyperplane arrangement the input belongs to.
For this reason, $I(\Yvh^N; \Xvh^N)$  is maximized by choosing an input support as the subset of {\small $\overline{\Nbb}^m$} in which a single point is contained in each partition  induced by $\{\Vv \Hv \xv=\tv \}$ and letting the input distribution be uniformly distributed over this set. 
As a final step of the proof, we note that the upper bound in \eqref{eq:tot outer} can be minimized over the choice of the set {\small $\overline{\Nbb}^m$} in \eqref{eq:nbb}. 
In other words, by varying the choice of $\nvb_{\nv}$ in \eqref{eq:nbb}, the points in {\small $\overline{\Nbb}^m$} are moved outside the corresponding partition, thus tightening the bound in \eqref{eq:tot outer}.
Accordingly, unless a partition contains a unit ball centered around a value $\nv \in \Rbb^{\Nrx}$, a value $\nvb_{\nv}$ can be chosen so that $\overline{\Nbb}^m$ does not contain a value in such partition. 
It then follows  that $ I(\Yvh^N; \Xvh^N) \leq  \log \rssps(\Acal,\sqrt{P})$ which is the desired result. 
\end{IEEEproof}
\begin{rem} Th. \ref{thm:main} extends the results in Sec. \ref{sec:Prior Results} as it holds for any set of possible receiver configurations $\Fcal$ in
	\eqref{eq:available conf}. The results in Sec. \ref{sec:Prior Results} only hold when $\Fcal$ has a specific form as in \eqref{eq:fcal a} or \eqref{eq:fcal b}.
On the other hand, Th. \ref{thm:main} does not provide a closed-form characterization of capacity as it involves the solution of a packing problem.
In particular, letting $\Fcal$ in \eqref{eq:Acal} have the form of \eqref{eq:fcal a} or \eqref{eq:fcal b} does not immediately recover the capacity characterization in Sec. \ref{sec:Prior Results} as Th. \ref{thm:main} follows a different approach than \cite{rini2017generalITW} to bound capacity. 
\end{rem}

\begin{rem}
When considering the model with any $\Hv$ and $\Vv$, the result in Th. \ref{thm:main} generalizes as follows.
The channel model in \eqref{eq:quantized model} is reduced to model where  $ \Vv$ and $ \Hv$ are such that $\diag\{\Hv \Hv^T\}=\diag\{\Vv \Vv^T\}=\ones_{1 \times \Nrx} $ by letting the additive noise $\Zv_n$ have a general covariance matrix.
For a channel model under this normalization, consider the additive noise after combining,  $\Zvt_{n}=\Vv \Zv_n$:
the variance of the $i^{\rm th}$ entry of $\Zvt_{n}$, $\Zt_{i,n}$, determines the uncertainty in the output of the $i^{\rm th}$ quantizer, $Y_{i,n}$.
Accordingly, the capacity is approximatively equal to the number of  separable points which can be fitted in the sphere of radius $\sqrt{P}$ such that 
each point is at distance at least $(\var[\Zt_{i,n}])^{1/2}$ from the $i^{\rm th}$ hyperplane. 
The complete derivation can be found in \cite{UStranDS}.
\end{rem}

\section{Conclusion}
\label{sec:Conclusion}
In this paper, the capacity of a MIMO channel with output quantization constraints for receivers equipped with analog combiners and one-bit threshold quantizers is investigated. 
The connection between the capacity of the system and a constrained sphere packing problem is showed by arguing that the threshold quantizers can be interpreted as hyperplanes partitioning the transmit signal space. 
This connection reveals, for example, that the infinite SNR capacity of a channel with linear combiner is attained by a receiver configuration which partitions the transmit signal space in the largest number of regions.

\bibliographystyle{IEEEtran}

\end{document}